\documentclass[11pt,a4paper]{scrartcl}

\usepackage{CLICdp}

\usepackage{CLICdp_definitions}

\usepackage{subcaption}

\renewcommand{\whizard}{\textsc{Whizard}\xspace}
\renewcommand{\pythia}{\textsc{Pythia}\xspace}
\newcommand{\delphes}{\textsc{Delphes}\xspace}

\newcommand{\cliclabel}[2]
           {
             
             \vspace*{-#1}
             \hspace*{#2}
{\fontencoding{T1}\fontfamily{phv}\fontseries{b}\selectfont
  \scriptsize\hspace*{-1cm}  
  CLICdp}
\vspace{-\baselineskip}
\vspace*{#1}
           }

\title{Sensitivity to
  invisible Higgs boson decays at CLIC}

\nolicence 

\clicdppub{2020}{002}  

\date{\today}

\addauthor{K.~Mekala}{}
\addauthor{A.~F.~Zarnecki}{}
\addauthor{B.~Grzadkowski}{}
\addauthor{M.~Iglicki}{}

\addinstitute{0}{Faculty of Physics, University of Warsaw, Poland}

\abstract{We studied the possibility of measuring invisible Higgs
  boson decays at CLIC running at 380\,GeV and 1.5\,TeV. 
  The analysis is based on the \whizard event generation and fast
  simulation of the CLIC detector response with \delphes. 
  We considered $\Pep\Pem$ background processes but also relevant
  $\PGg\PGg$ and $\PGg\Pe^{\pm}$ interactions. The approach
  consisting of a two step analysis was used to optimize separation
  between signal and background processes. First, a set of
  preselection cuts was applied; then, multivariate analysis
  methods were employed to optimise the significance of observations. We
  estimated the expected limits on the invisible decays of the 125\,GeV
  Higgs boson, as well as the cross section limits for production of
  an additional neutral Higgs-like scalar, assuming its invisible
  decays, as a function of its mass. Extracted model-independent
  branching ratio and cross section limits were then interpreted in
  the framework of the vector-fermion dark matter model to set limits
  on the mixing angle between the SM-like Higss boson and the new
  scalar of the "dark sector".} 

\titlecomment{This work was carried out in the framework of the CLICdp
              Collaboration} 

\graphicspath{ {./logos/}{./figures/} }

\begin{document}


\titlepage

\section{Introduction}
All available experimental results seem to confirm that the new
particle discovered in 2012 by the \mbox{ATLAS} and CMS experiments at
LHC \cite{odH, odcms} is the last missing constituent of the Standard
Model (SM), the Higgs boson.
The Standard Model predicts that the Higgs boson with a mass of about
125\,GeV should decay predominantly to $b\bar{b}$ (about 58\% of all
decays), but also to $WW^*$ (21\%), $\tau\bar{\tau}$ (6,3\%) or
$ZZ^*$ (2,6\%)\cite{pdg}. 
Some extensions of the Standard Model predict "invisible" Higgs boson
decays -- into new, not-observable particles.
These particles could contribute to the Dark Matter (DM) density of the
Universe.
As of today, the best limits on invisible Higgs boson decays come
from the CMS experiment -- at 95\% CL the branching fraction is less
than 19\%\cite{CMS}. 
We study prospects for constraining invisible decays of the 125 GeV
Higgs boson  at a future experiment at CLIC.

Among many theories introducing DM particles and predicting invisible
decays of the SM-like Higgs boson, there is a group of "Higgs-portal"
models which assume existence of additional fundamental scalars.
These new particles could mix with the SM-like Higgs and thus open new
decay channels of the SM-like 125 GeV state into DM particles.
The new scalars, if they are relatively light, could also be produced
in $\Pep\Pem$ collisions in the process corresponding to the
Higgstrahlung process in the SM: $\Pep\Pem \to \PZ\PH'$.
We extend our search for invisible decays of the SM-like Higgs boson
to the search for production and invisible decays of a scalar
particle, $\PH'$, with arbitrary mass.
We then interpret our results in terms of the limits on the scalar
sector mixing angle.
We considered the \emph{Vector-fermion dark matter} model
(VFDM)\cite{vfdm1,vfdm2}, a simple extension of the Standard Model with one extra scalar, two Majorana fermions and one gauge boson.

\section{Event generation and fast simulation of the detector response}

Results presented in this paper are based on a realistic simulation
of events, including fast simulation of the CLICdet \cite{clicdet}
detector response.
Signal and background event samples were generated using \whizard 2.7.0
\cite{whiz1, whiz2}, taking into account the beam energy profile
expected for CLIC running at 380\,GeV and 1.5\,TeV.
As the signal, we considered the Higgs-strahlung process: Higgs boson
production (with decay into an invisible final state) together with a
$\PZ$ boson decaying into a quark-antiquark pair.
For the background, we studied processes both with and without 
Higgs boson production.
We also took into account the possible background contribution from
$\PGg\PGg$ and $\Pepm\PGg$ interactions, where both beamstrahlung (BS)
photons and photon radiation by the incoming electrons, as described
by the Effective Photon Approximation (EPA)\footnote{For generation of EPA
events \whizard 2.8.3 was used, with a fix introduced to give correct 
matching of $\Pep\Pem$ and $\PGg^{EPA}\Pepm$ samples.}, 
were taken into account.
For CLIC running at 380\,GeV, the same integrated luminosity of data is
expected to be taken with negative and positive electron beam
polarisation. We assume combined analysis of both
samples, corresponding to 1000\,fb$^{-1}$ collected with unpolarised
beam. At 1.5\,TeV we consider the two electron beam polarisations separately,
assuming 2000\,fb$^{-1}$ to be collected with -80\% polarisation and
500\,fb$^{-1}$ with +80\% polarisation\footnote{For CLIC running at 1.5\,TeV 
the integrated luminosity for $\PGg^{BS}\PGg^{BS}$ interactions is 
assumed to be about 64\% of the $\Pep\Pem$ luminosity and for $\Pep\PGg^{BS}$
and $\PGg^{BS}\Pem$ interactions -- about 75\%.} \cite{gHZZ}.
Cross sections for processes taken into account in the presented
study\footnote{Not included are $\Pem\Pep$ processes, 
  in particular those with six fermions
  in the final state, eg. $\PQq\PQq\PQq\PQq\Pl\PGn$ and
  $\PQq\PQq\PQq\PQq\PGn\PGn$, for which no events passed preselection cuts.
  For interactions involving EPA photons, only the process with the largest
  background countribution, $\PGg\Pepm\to\PQq\PQq\PGn$, was considered,
  as $\PGg\PGg$ and $\Pepm\PGg$ interactions are dominated by BS photon
  interactions.}
calculated by \whizard\footnote{Statistical uncertainty of
  cross-sections was below 1\% and has been neglected.} and numbers
  of generated events are shown in Tables~\ref{tab1} (for 380\,GeV
  running) and \ref{tab2} (for 1.5\,TeV).

\begin{table}[htbp]
\begin{center}
\begin{tabular}{|c|c|c|}
\hline
\textbf{\large Final state} & \textbf{\large $\sigma$ [fb]} & \textbf{\large N$_{GEN}$} \\ \hline
$\PQq\PQq$ & 22145.90 & 2000000 \\ 
$\Pl\Pl$ & 19917.10 & 1000000 \\ 
$\PQq\PQq\PQq\PQq$ & 5075.88 & 500000 \\ 
$\PQq\PQq\Pl\Pl$ & 1718.07 & 200000 \\ 
$\PQq\PQq\PGn\PGn$ & 317.44 & 300000 \\ 
$\PQq\PQq\Pl\PGn$ & 5557.10 & 500000 \\ 
$\PQq\PQq\Pl\PGn\PGn\PGn$ & 1.37 & 100000 \\ \hline
$\PH_{SM} + \PQq\PQq$ & 82.23 & 100000 \\ 
$\PH_{SM} + \Pl\Pl$ & 15.47 & 100000 \\ 
$\PH_{SM} + \PGn\PGn$ & 54.54 & 100000 \\  \hline
$\PGg^{BS}\PGg^{BS} \to \PQq\PQq$ & 1914.43 & 200000 \\ 
$\PGg^{BS}\PGg^{BS} \to \PQq\PQq\PQq\PQq$ & 1.84 & 10000 \\
$\PGg^{BS}\PGg^{BS} \to \PQq\PQq\Pl\Pl$ & 33.04 & 10000 \\ 
$\PGg^{BS}\PGg^{BS} \to \PQq\PQq\Pl\PGn$ & 0.72 & 10000 \\ 
$\PGg^{BS}\PGg^{BS} \to \PQq\PQq\PGn\PGn$ & 0.03 & 10000 \\ \hline
$\PGg^{BS}\Pem \to \PQq\PQq\PGn$ & 1418.31 & 300000 \\
$\PGg^{BS}\Pep \to \PQq\PQq\PGn$ & 1428.57 & 300000 \\ \hline
$\PGg^{EPA}\Pem \to \PQq\PQq\PGn$ & 883.29 & 100000 \\
$\PGg^{EPA}\Pep \to \PQq\PQq\PGn$ & 883.41 & 100000 \\ \hline
$\PH_{inv} + \PQq\PQq$ & & 100000 \\ \hline
\end{tabular}
\end{center}
\caption{Cross sections, $\sigma$, and numbers of generated events,
  N$_{GEN}$, for each final state considered at 380\,GeV; the
  reference cross section for $\PH_{inv}\PQq\PQq$ channel is the SM
  Higgs boson production cross section for $\PH_{SM}\PQq\PQq$ final
  state.} 
\label{tab1}
\end{table}

\begin{table}[htbp]
\begin{center}
\begin{tabular}{|c|c|c|c|}
\hline
\textbf{\large Final state} & \textbf{\large $\sigma^{neg}$ [fb]} &
   \textbf{\large $\sigma^{pos}$ [fb]} & \textbf{\large N$_{GEN}$} \\ \hline
$\PQq\PQq$ & 2873.36 & 1807.09 & 1000000 \\ 
$\Pl\Pl$ & 1396.18 & 1218.17 & 1000000 \\ 
$\PQq\PQq\PQq\PQq$ & 1970.67 & 265.47 & 1000000 \\ 
$\PQq\PQq\Pl\Pl$ & 2739.19 & 2570.04 & 1000000 \\ 
$\PQq\PQq\PGn\PGn$ & 1520.11 & 187.32 & 1000000 \\ 
$\PQq\PQq\Pl\PGn$ & 7054.84 & 1712.63 & 1000000 \\ 
$\PQq\PQq\Pl\PGn\PGn\PGn$ & 40.15 & 5.39 & 100000 \\ \hline
$\PH_{SM} + \PQq\PQq$ & 9.42 & 6.59 & 100000 \\ 
$\PH_{SM} + \Pl\Pl$ & 31.65 & 22.09 & 100000 \\ 
$\PH_{SM} + \PGn\PGn$ & 467.62 & 53.49 & 100000 \\  \hline
$\PGg^{BS}\PGg^{BS} \to \PQq\PQq$ & 6030.48 & 6030.48 & 1000000 \\ 
$\PGg^{BS}\PGg^{BS} \to \PQq\PQq\PQq\PQq$ & 6330.76 & 6330.76 & 100000 \\
$\PGg^{BS}\PGg^{BS} \to \PQq\PQq\Pl\Pl$ & 2791.48 & 2791.48 & 100000 \\ 
$\PGg^{BS}\PGg^{BS} \to \PQq\PQq\Pl\PGn$ & 6697.38 & 6697.38 & 100000 \\ 
$\PGg^{BS}\PGg^{BS} \to \PQq\PQq\PGn\PGn$ & 6.57 & 6.57 & 100000 \\ \hline
$\PGg^{BS}\Pem \to \PQq\PQq\PGn$ & 28274.29 & 3141.6 & 1000000(neg.)/100000(pos.)\\
$\PGg^{BS}\Pep \to \PQq\PQq\PGn$ & 15621.69 & 15621.69 & 1000000 \\ \hline
$\PGg^{EPA}\Pem \to \PQq\PQq\PGn$ & 6616.77 & 735.20 & 300000 \\
$\PGg^{EPA}\Pep \to \PQq\PQq\PGn$ & 3677.43 & 3677.43 & 300000 \\ \hline
$\PH_{inv} + \PQq\PQq$ &      &      & 100000 \\ \hline
\end{tabular}
\end{center}
\caption{Cross sections, $\sigma^{neg}$ and $\sigma^{pos}$, for -80\%
  and +80\% electron beam polarisation, respectively, at 1.5\,TeV CLIC,
  and numbers of generated events, N$_{GEN}$, for each considered 
  final state; the reference cross sections for $\PH_{inv}\PQq\PQq$
  channel are the SM Higgs boson production cross sections for
  $\PH_{SM}\PQq\PQq$ final state.} 
\label{tab2}
\end{table}

To simulate detector response the fast simulation framework \delphes
was used \cite{delph}. Control cards prepared for the new detector
model CLICdet \cite{delcards} were modified to make Higgs particles
`invisible' in the simulation (ignored when generating detector
response), so that the invisible Higgs boson decays can be modeled by
defining the Higgs boson as stable in \whizard and \pythia. 
An event reconstruction begins with searching for isolated electrons,
muons and photons (assuming reconstruction efficiency resulting from
the full detector simulation).
For CLICdet, \delphes identifies isolated electrons and photons with
energy of at least 2~GeV, and muons of 3~GeV.
Jet clustering was carried out with the VLC\cite{vlc} algorithm
assuming minimal transverse jet momentum of 20~GeV. The values of $R=1.5$
for 380\,GeV and $R=0.7$ for 1.5\,TeV CLIC running were chosen as
optimal for the VLC algorithm, with $\beta$ = $\gamma$ = 1.
We require reconstruction of two hadronic jets, each with at least two
charged particles.
To take into account the effects of the beam-induced backgrounds,
additional energy smearing is applied for jets reconstructed
at 1.5\,TeV:
for central jets ($|\eta|<0.76$) the overlay events are expected
to result in an additional 1\% jet energy smearing, while for more
forward jets ($|\eta|\ge0.76$) 5\% smearing is assumed
\cite{delcards}.\footnote{Jet energy smearing expected for CLIC runing
 at 1.5\,TeV (stage 2) and 3\,TeV (stage 3) is implemented in \delphes 
 cards for CLICdet \cite{delcards}. However,
  only the jet energy is smeared and the jet mass is assumed to be
  negligible. As the jet masses should not be neglected in the
  presented analysis, we use our own implementation of the jet energy
  smearing, with energy and mass of the jet scaled by the same factor.}   

\section{Signal event selection}

\subsection{Preselection of events}
A main purpose of the preselection is to remove all background events
which are not consistent with the expected signature of the signal
process.
For the process $\Pep\Pem\to \PH\PZ \to inv+\PQq\PQq$, we expect to
observe only two jets in the final state, with an invariant mass
consistent with the mass of the Z boson.
In the initial preselection step, all events which were not consistent
with this signature were rejected.
In particular, events with isolated leptons (electrons or muons) or
isolated energetic photons (with energy greater than 5~GeV) were
excluded.
For a significant fraction of events, the difference between the
energy sum of the reconstructed jets and the energy sum of all
identified particles in the event was sizable, indicating an
incomplete event reconstruction.
To avoid such events, we require that this difference is less than
10~GeV. 

\begin{figure}[tbp]
  \begin{subfigure}[b]{0.48\textwidth}
    \includegraphics[width=\textwidth]{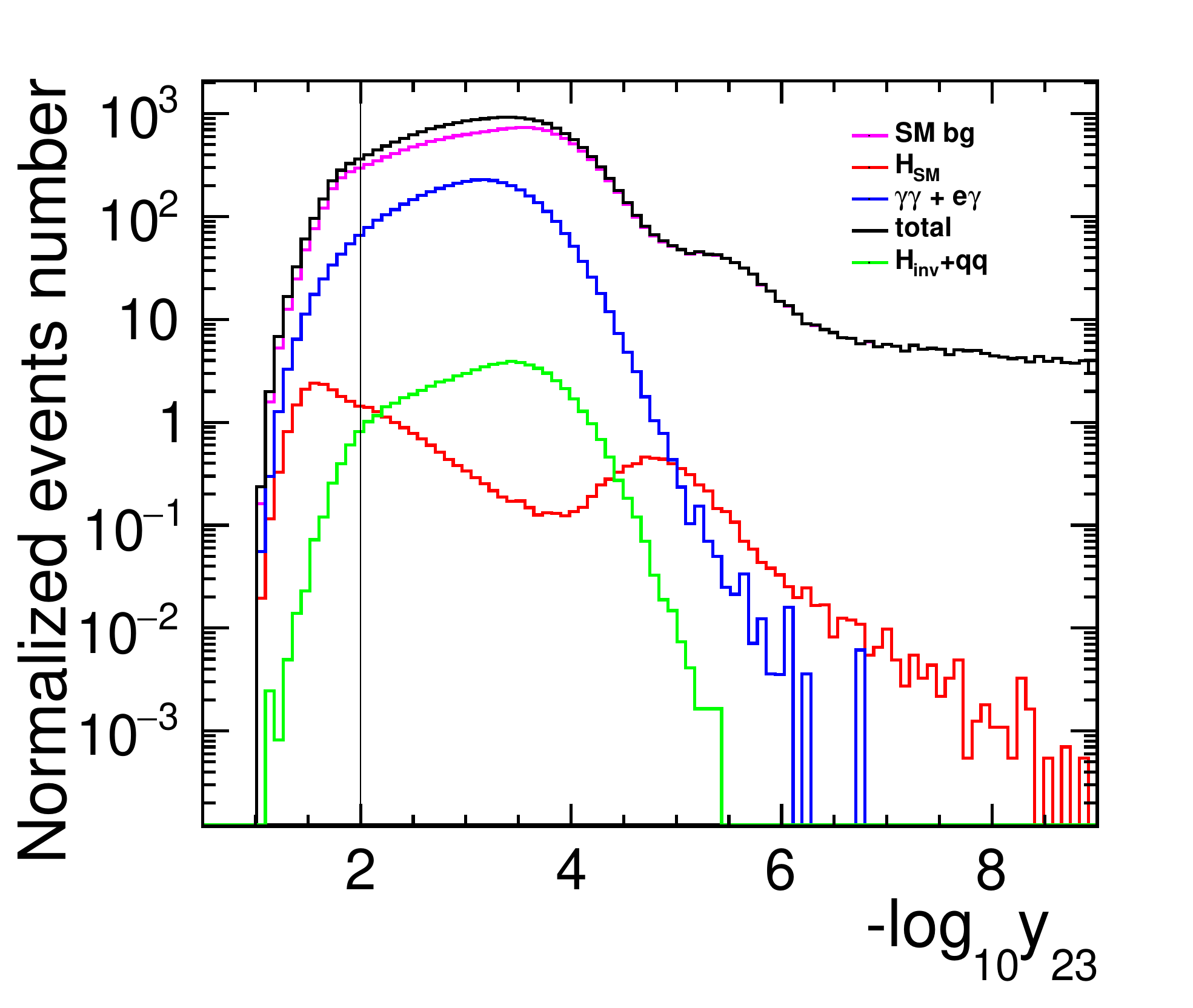}
    \caption{The jet clustering parameter y$_{23}$.}
    \label{fig:y23}
  \end{subfigure}%
  \hfill 
  \begin{subfigure}[b]{0.48\textwidth}
    \includegraphics[width=\textwidth]{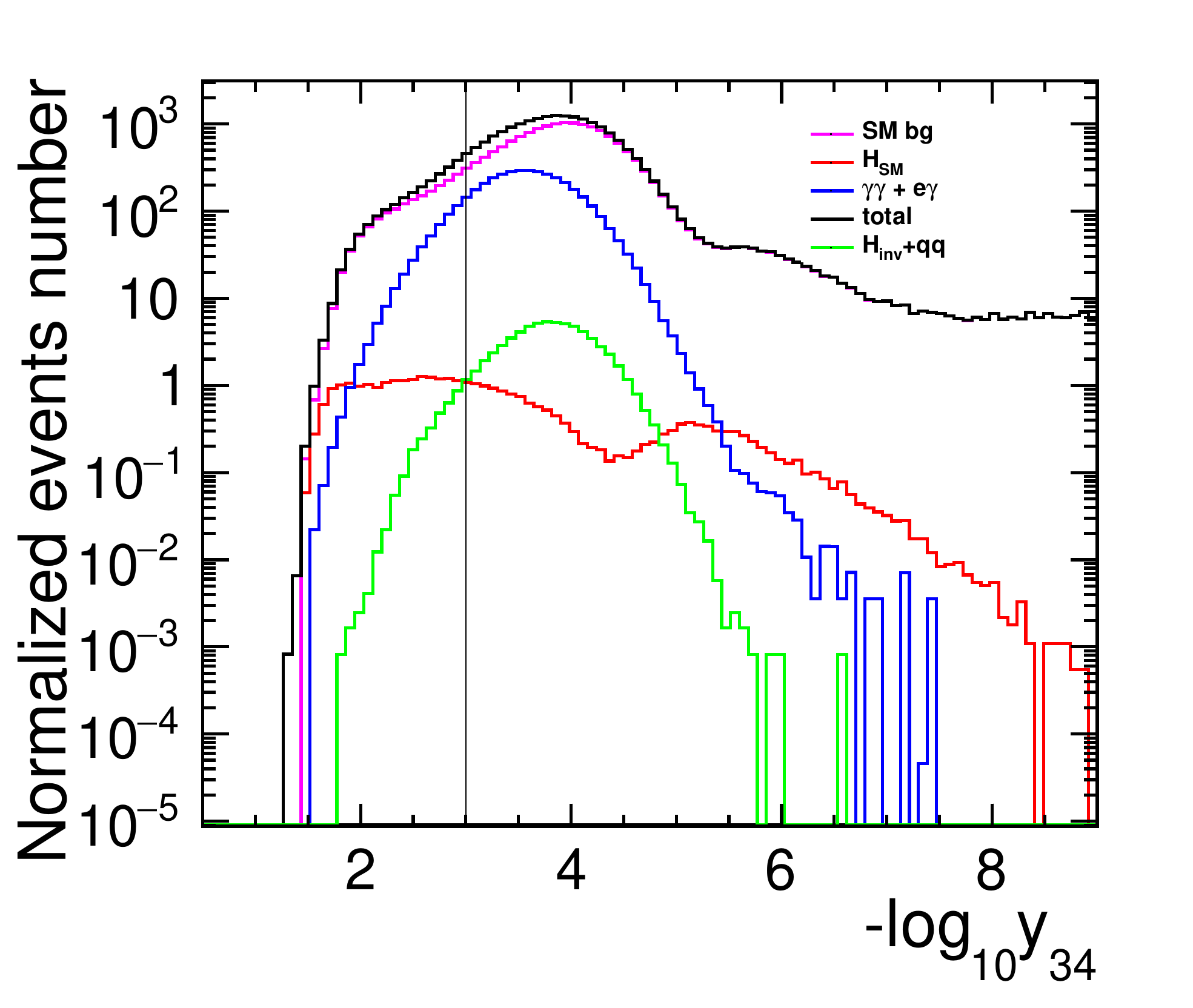}
    \caption{The jet clustering parameter y$_{34}$.}
    \label{fig:y34}
  \end{subfigure}
  \begin{subfigure}[b]{0.48\textwidth}
  \centering
    \includegraphics[width=\textwidth]{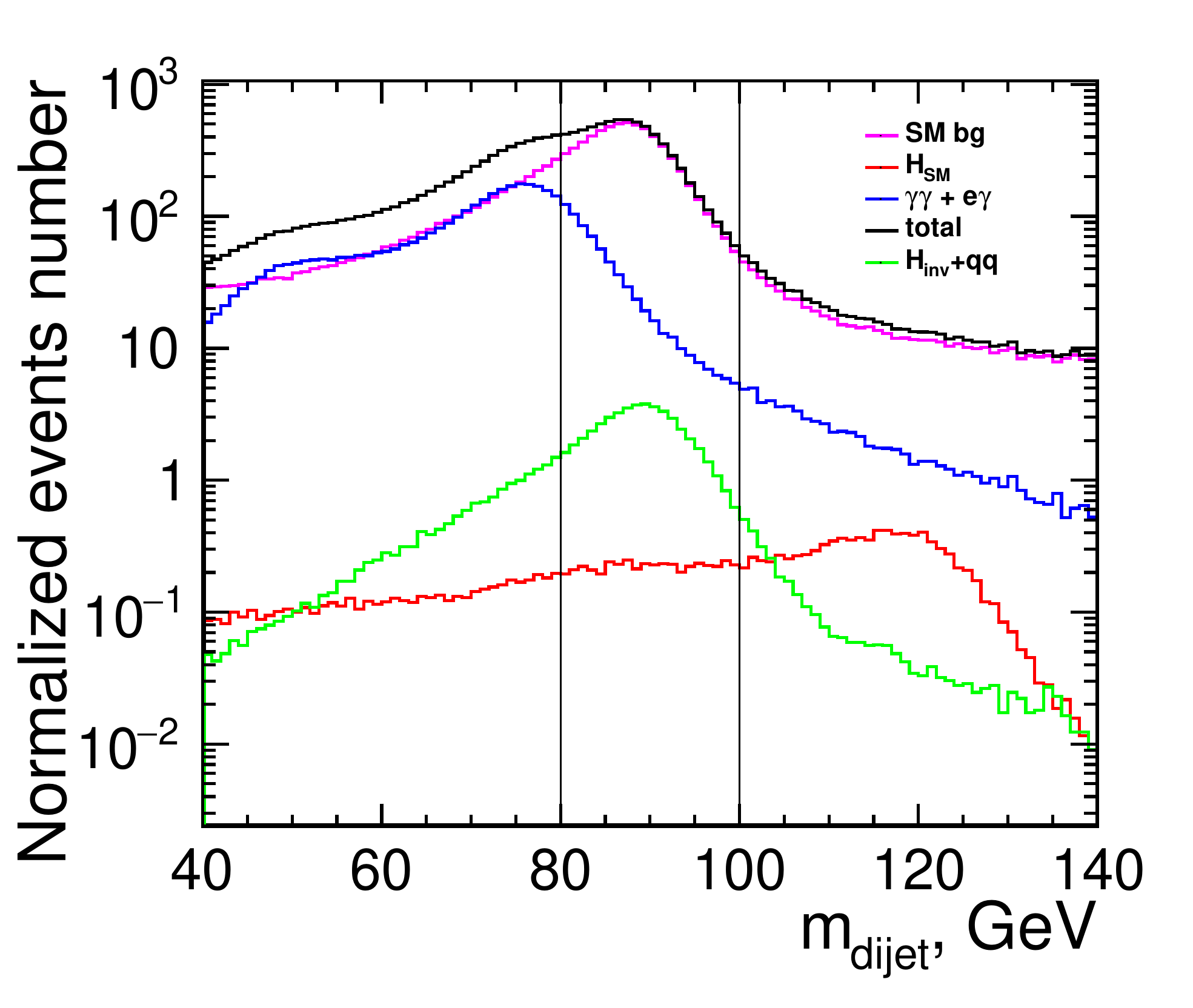}
    \caption{The dijet invariant mass.}
    \label{fig:dijet}
  \end{subfigure}%
  \hfill 
  \begin{subfigure}[b]{0.48\textwidth}
  \centering
    \includegraphics[width=\textwidth]{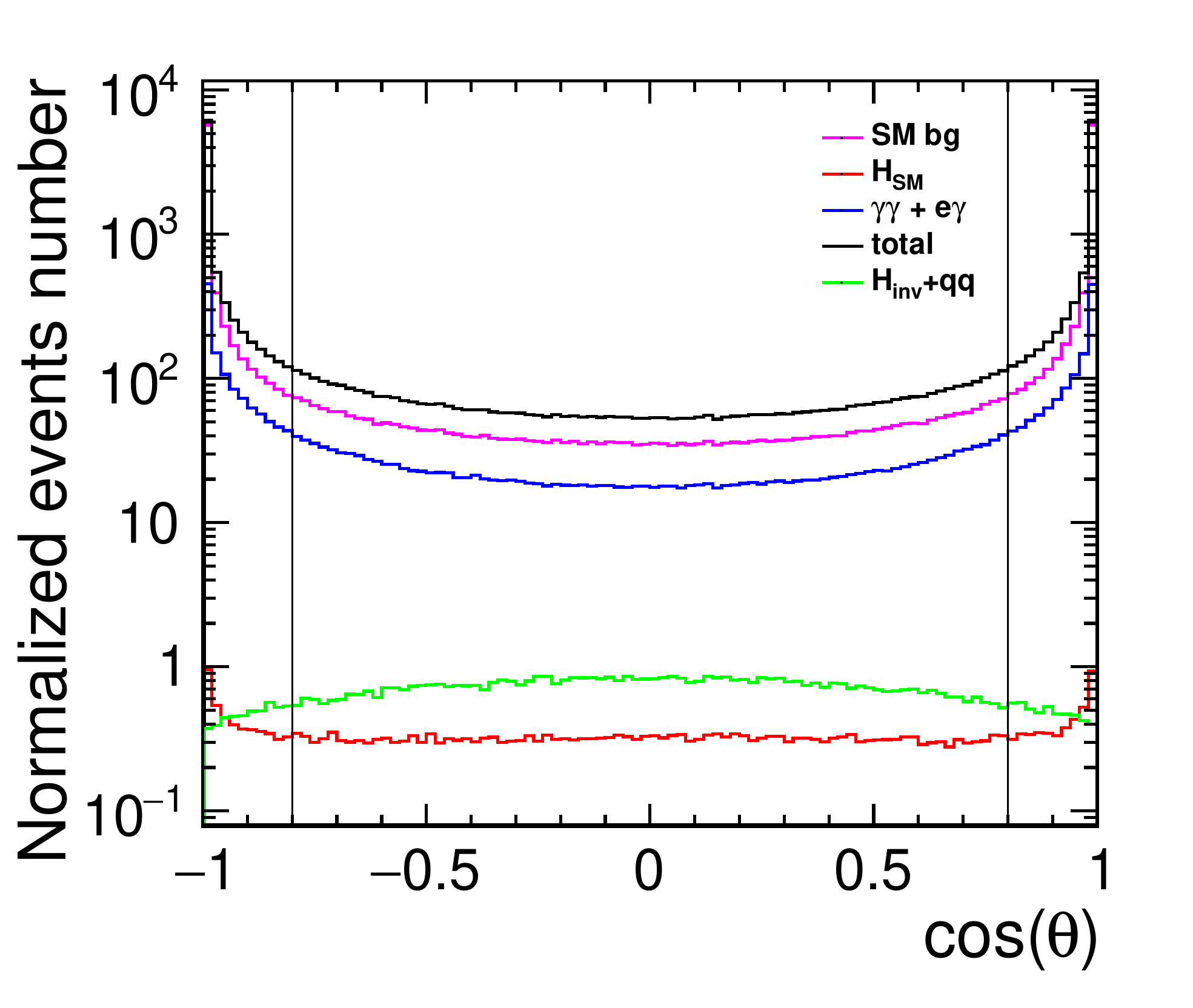}
    \caption{The cosine of the dijet emission angle.}
    \label{fig:theta}
  \end{subfigure}
    \cliclabel{14.3cm}{2.2cm}
    \cliclabel{14.3cm}{10.5cm}
    \cliclabel{7.15cm}{2.2cm}
    \cliclabel{7.15cm}{10.5cm}
  \caption{Distribution of preselection variables for different event
    samples considered: total background (black), SM background without Higgs particle
    production (pink), background of SM Higgs boson production and decays (red), photon interactions (blue) and
    signal (green); black vertical lines indicates preselection cuts,
    see text for details.} 
\end{figure}

In the next steps, quantities describing event topology were
considered. 
First, we analyzed the distributions of parameters y$_{23}$ and
y$_{34}$ describing the results of jet clustering with VLC algorithm.
While the algorithm was forced to reconstruct two jets in each
event, these distributions allowed us to distinguish actual two-jet
events from events with a larger number of underlying jets in the final
state.
The distributions of -log$_{10}$y$_{23}$ and -log$_{10}$y$_{34}$ are
shown in Figures \ref{fig:y23} and \ref{fig:y34}, respectively.
The double-peak structure clearly visible in the background sample of
SM Higgs boson decays $\PH_{SM}$ corresponds to pure two- and
four-jet events.
The distribution for the other background channels is more uniform.
For the signal sample, with two hadronic jets, we expect values of
y$_{23}$ and y$_{34}$ to be relatively small.
We selected events for which y$_{23}<0.01 $ (-log$_{10}$y$_{23}>$ 2.0)
and y$_{34}<0.001$ (-log$_{10}$y$_{34}>$3.0).

The next quantity considered for the preselection of signal events was
the invariant mass of the two jet final state -- m$_{dijet}$.
It should correspond to the mass of the $\PZ$ boson, so only events
for which this value was in the range of 80-100~GeV were selected for
further analysis. 
The distribution of invariant masses for each channel is shown in the
Figure \ref{fig:dijet}.
The peak visible in the channel of SM Higgs boson decays ($\PH_{SM}$)
around 120~GeV corresponds to the process $\Pep\Pem\to \PH\PZ \to
\PQq\PQq\PGn\PGn$ which, considering only its topology (two jets and
missing energy), is indistinguishable from the signal.
We also studied the distribution of the dijet emission angle,
$\theta$, defined as the angle between the beam axis and a sum of the
jet four-momenta (the emission angle of the \PZ boson for the signal events).
For the majority of background events small emission angles are reconstructed,
while for signal events the distribution is almost flat (angles close
to 90$^\circ$ are slightly preferred).
Therefor, events for which $|\cos(\theta)|$ was greater than 0.8 were
excluded from the analysis.
The distribution of cosine of the angle $\theta $ is shown in the
Figure \ref{fig:theta}.

\begin{figure}[htbp]
  \centerline{\includegraphics[width=0.6\textwidth]{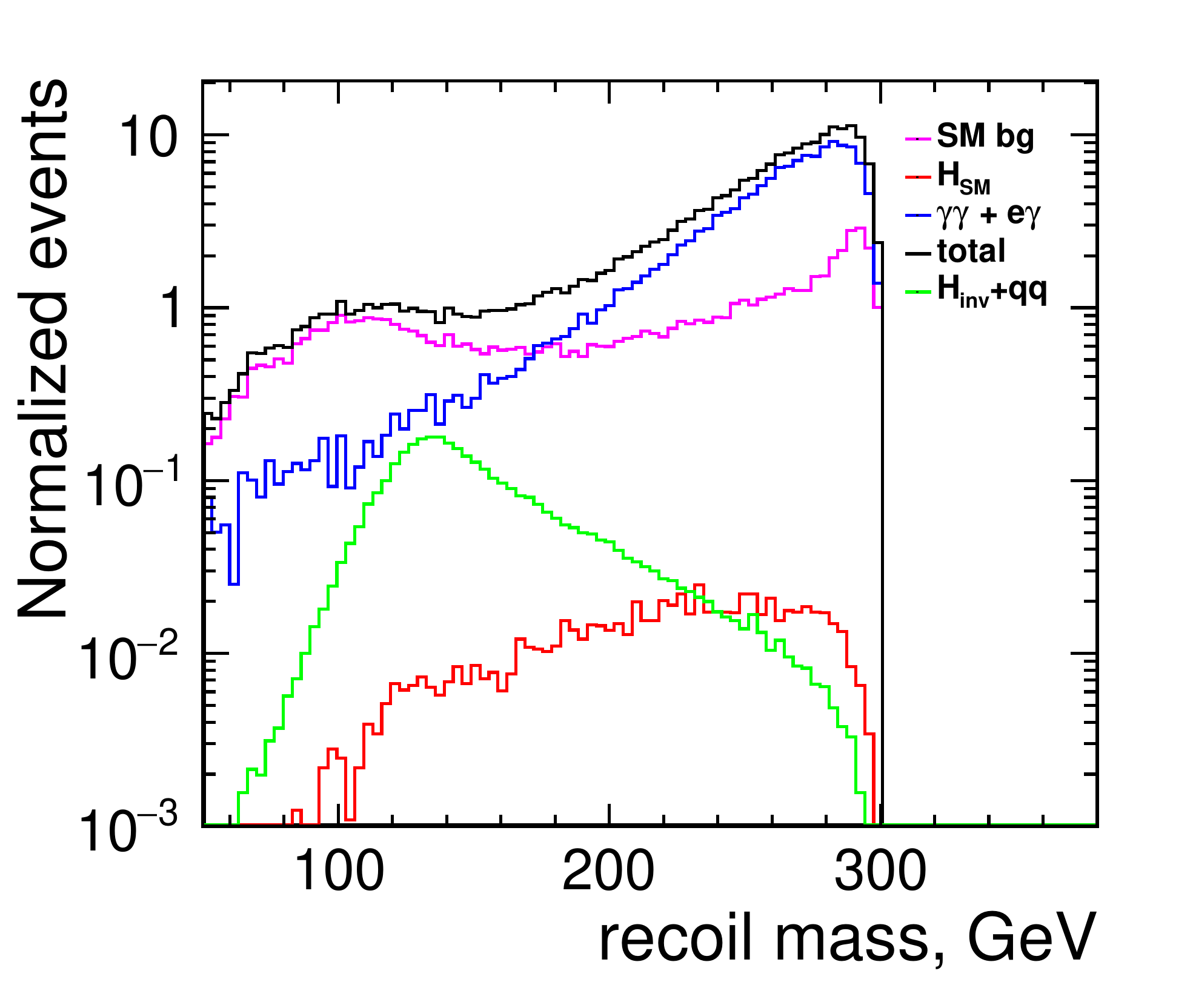}}
    \cliclabel{8cm}{5.7cm}
  \caption{Distributions of the reconstructed invariant mass of the
    invisible Higgs boson decay products (recoil mass) expected for
    background (total background -- black, SM background without Higgs particle production -- pink, background of SM Higgs boson production and decays -- red, photon interactions -- blue) and signal (green) events after preselection cuts, assuming
    integrated $\Pep\Pem$ luminosity of 1000 fb$^{-1}$ collected at 380\,GeV and
    BR$(\PH \to inv)=10\%$ for signal events.
}
  \label{fig:mrec_sm}
\end{figure}

The results of the preselection are presented in Tables \ref{tab3}
and \ref{tab4}.
Shown in Fig.~\ref{fig:mrec_sm} is the expected distribution of the so
called recoil mass, the invariant mass of the Higgs boson produced
together with the Z boson, after preselection cuts, reconstructed from
the energy-momentum conservation for CLIC running at 380\,GeV.
For the background sample the distribution has two maxima: at around
300\,GeV, which is the maximum recoil mass allowed (as we require two
jets to have an invariant mass of at least 80\,GeV) and at around
90\,GeV, which is mainly due to invisible Z boson decays.
For signal events, normalised in Fig.~\ref{fig:mrec_sm} to BR$(\PH \to
inv)=10\%$, the expected recoil mass distribution is consistent with
the SM Higgs boson mass of 125\,GeV.
The slight shift of the maxima in the reconstructed recoil mass
distributions towards higher mass values is most likely
due to the influence of the beam energy spectra (a significant fraction
of events takes place at energy scales lower than the nominal one). 

\begin{table}[btp]
\begin{center}
\begin{tabular}{|c|c|c|}
\hline
\textbf{Final state} & \textbf{Efficiency} & \textbf{\large N$_{pre}$} \\ \hline
\multicolumn{3}{|c|}{without Higgs boson} \\ \hline
$\PQq\PQq$ & 0.08\% & 16831\\ 
$\Pl\Pl$ & <0.01\% & 20 \\ 
$\PQq\PQq\PQq\PQq$ & <0.01\% & 51 \\ 
$\PQq\PQq\Pl\Pl$ & 0.02\% & 412 \\ 
$\PQq\PQq\nu\nu$ & 20.47\% & 64974 \\ 
$\PQq\PQq\Pl\nu$ & 0.60\% & 33598 \\ 
$\PQq\PQq\Pl\nu\nu\nu$ & 1.32\% & 18 \\ 
\textbf{Total:} & 0.21\% & 115904 \\ \hline
\multicolumn{3}{|c|}{with Higgs boson decays described in the Standard Model} \\ \hline
$\PH_{SM}+\PQq\PQq$ & <0.01\% & 41 \\  
$\PH_{SM}+\Pl\Pl$ & 0.01\% & 1 \\  
$\PH_{SM}+\nu\nu$ & 2.33\% & 1273 \\  
\textbf{Total:} & 0.86\% & 1315 \\ \hline
\multicolumn{3}{|c|}{photon interactions} \\ \hline
$\PGg^{BS}\PGg^{BS} \to \PQq\PQq$ & 0.78\% & 9496 \\ 
$\PGg^{BS}\PGg^{BS} \to \PQq\PQq\Pl\Pl$ & 0.03\% & 6 \\ 
$\PGg^{BS}\PGg^{BS} \to \PQq\PQq\Pl\PGn$ & 1.59\% & 7 \\ 
$\PGg^{BS}\PGg^{BS} \to \PQq\PQq\PGn\PGn$ & 0.06\% & 1 \\
$\PGg^{BS}\PGg^{BS} \to \PQq\PQq\PQq\PQq$ & 0.03\% & 1 \\
$\PGg^{BS}\Pem \to \PQq\PQq\PGn$ & 6.73\% & 71596 \\
$\PGg^{BS}\Pep \to \PQq\PQq\PGn$ & 6.68\% & 71561 \\
$\PGg^{EPA}\Pem \to \PQq\PQq\PGn$ & 6.32\% & 55833 \\
$\PGg^{EPA}\Pep \to \PQq\PQq\PGn$ & 6.20\% & 54771 \\
\textbf{Total:} & 5.11\% & 263270 \\ \hline
\multicolumn{3}{|c|}{signal} \\ \hline
$\PH_{inv}+\PQq\PQq$ & 43.56\% & 35779 \\ \hline
\end{tabular}
\caption{Efficiency and expected events number after preselection
  N$_{pre}$ assuming integrated $\Pep\Pem$ luminosity of 1000 fb$^{-1}$ collected
  at 380\,GeV for each considered final state. Results for the signal
  were calculated assuming SM Higgs boson production cross section and
  BR$(\PH \to inv)=100\%$.} 
\label{tab3}
\end{center}
\end{table}
\begin{table}[btp]
\begin{center}
\begin{tabular}{|c|c|c|c|c|}
\hline
\textbf{Final state} & \textbf{Efficiency - p. neg.} & \textbf{\large N$^{neg}_{pre}$}  & \textbf{Efficiency - p. pos.} & \textbf{\large N$^{pos}_{pre}$} \\ \hline
\multicolumn{5}{|c|}{without Higgs boson} \\ \hline
$\PQq\PQq$ & 0.07\% & 4040 & 0.08\% & 685\\  
$\Pl\Pl$ & <0.01\% & 6 & <0.01\% & 2 \\ 
$\PQq\PQq\PQq\PQq$ & <0.01\% & 118 & <0.01\% & 7 \\ 
$\PQq\PQq\Pl\Pl$ & 0.04\% & 2180 & 0.04\% & 458 \\ 
$\PQq\PQq\nu\nu$ & 13.55\% & 412029 & 12.73\% & 11920 \\ 
$\PQq\PQq\Pl\nu$ & 1.47\% & 207723 & 2.29\% & 19622 \\ 
$\PQq\PQq\Pl\nu\nu\nu$ & 1.25\% & 1001 & 2.10\% & 57 \\ 
\textbf{Total:} & 1.76\% & 627097 & 0.84\% & 32751 \\  \hline
\multicolumn{5}{|c|}{with Higgs boson decays described in the Standard Model} \\ \hline
$\PH_{SM}+\PQq\PQq$ & 0.02\% & 3 & <0.01\% & 0 \\ 
$\PH_{SM}+\Pl\Pl$ & 0.11\% & 69 & 0.11\% & 12 \\ 
$\PH_{SM}+\nu\nu$ & 2.34\% & 21903 & 2.50\% & 667 \\ 
\textbf{Total:} & 2.16\% & 21976 & 1.65\% & 680 \\ \hline
\multicolumn{5}{|c|}{photon interactions} \\ \hline
$\PGg^{BS}\PGg^{BS} \to \PQq\PQq$ & 0.21\% & 16541 & 0.22\% & 4221 \\ 
$\PGg^{BS}\PGg^{BS} \to \PQq\PQq\Pl\Pl$ & 0.04\% & 1501 & 0.04\% & 375 \\ 
$\PGg^{BS}\PGg^{BS} \to \PQq\PQq\Pl\PGn$ & 0.32\% & 27518 & 0.32\% & 6815 \\ 
$\PGg^{BS}\PGg^{BS} \to \PQq\PQq\PGn\PGn$ & 0.38\% & 32 & 0.37\% & 8 \\
$\PGg^{BS}\PGg^{BS} \to \PQq\PQq\PQq\PQq$ & <0.01\% & 0 & <0.01\% & 0 \\
$\PGg^{BS}\Pem \to \PQq\PQq\PGn$ & 4.04\% & 1711301 & 3.91\% & 46041 \\
$\PGg^{BS}\Pep \to \PQq\PQq\PGn$ & 4.02\% & 942574 & 4.01\% & 234665 \\
$\PGg^{EPA}\Pem \to \PQq\PQq\PGn$ & 3.15\% & 416636 & 3.16\% & 11610 \\
$\PGg^{EPA}\Pep \to \PQq\PQq\PGn$ & 3.22\% & 236532 & 3.23\% & 59335 \\
\textbf{Total:} & 2.93\% & 3352635 & 2.24\% & 363070 \\ \hline
\multicolumn{5}{|c|}{signal} \\ \hline
$H_{inv}$ + $qq$ & 42.16\% & 8023 & 42.04\% & 1388 \\ \hline
\end{tabular}
\caption{Efficiency and expected events number after preselection
  N$_{pre}$ assuming integrated $\Pep\Pem$ luminosity of 2000 fb$^{-1}$ 
  (negative polarisation) and 500 fb$^{-1}$ (positive polarisation)
  collected at 1.5\,GeV for each considered final state. Results 
  for the signal were calculated assuming SM Higgs boson 
  production cross section and  BR$(\PH \to inv)=100\%$.} 
\label{tab4}
\end{center}
\end{table}

\subsection{Final selection}

The second stage of the analysis was based on multivariate analysis
and machine learning.
The Boosted Decision Trees (BDT)\cite{tree} algorithm, as implemented in
TMVA framework\cite{TMVA}, was used, with 1000 trees and 5 input
variables.
The following parameters were selected as the BDT input variables:
\begin{enumerate}
\item E$_{j\!j}$ -- dijet energy,
\item m$_{j\!j}$ -- dijet invariant mass,
\item $\alpha_{j\!j}$ -- angle between the two reconstructed jets in the LAB frame,
\item m$^{miss}$ -- reconstructed missing mass,
\item p$^{miss}_{t}$ -- missing transverse momentum.
\end{enumerate}

\begin{figure}
  \includegraphics[width=\textwidth]{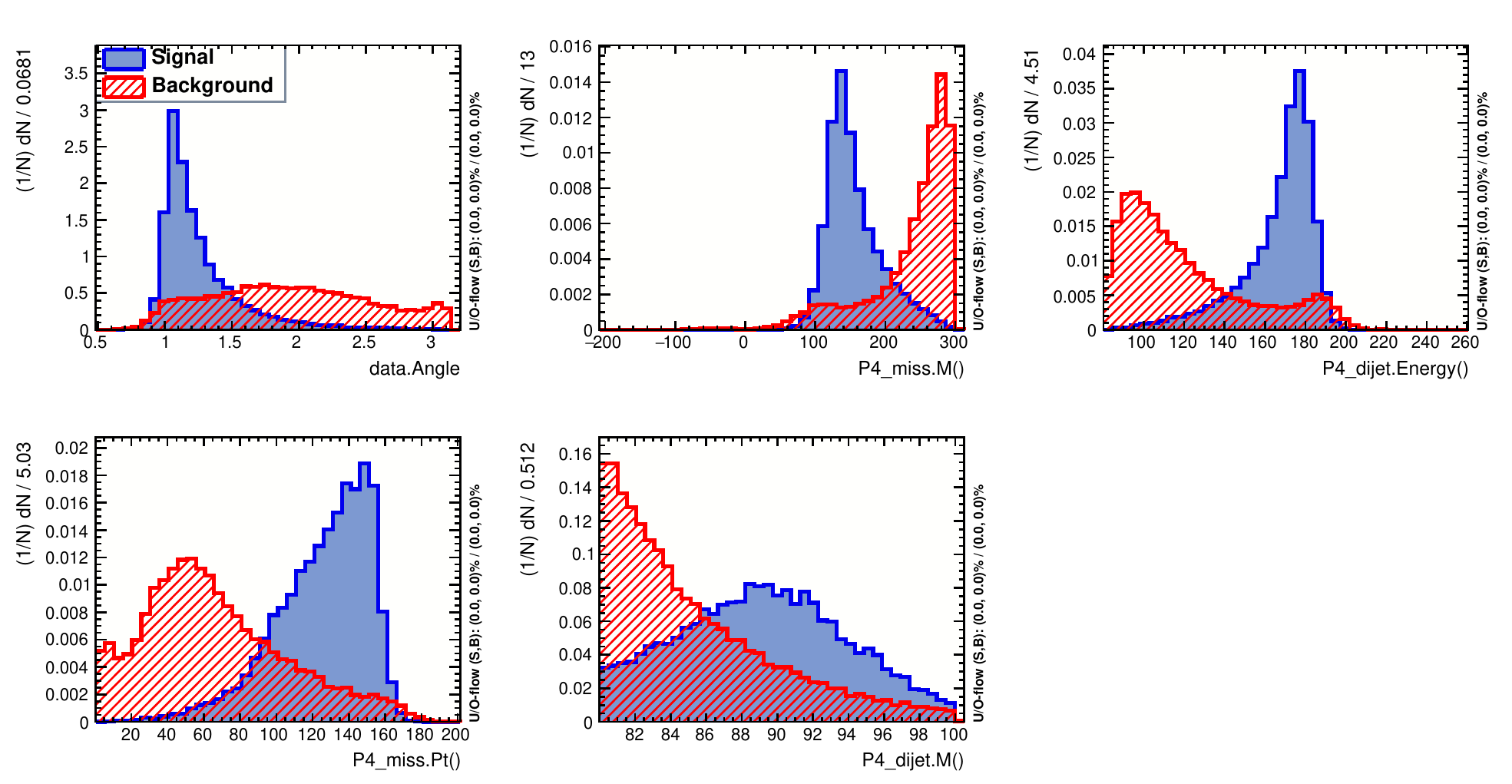}
  \cliclabel{8.9cm}{1.85cm}\vspace*{-5mm}
  \caption{Distributions of the input variables for the BDT algorithm,
    for signal (blue) and background (red) event samples, for CLIC
    running at 380\,GeV. Top row (from left): angle $\alpha_{jj}$, m$^{miss}$
    [GeV], E$_{jj}$ [GeV]; bottom row: p$_{t}^{miss}$ [GeV], m$_{jj}$
    [GeV].} 
  \label{fig:variables}
\end{figure}

\begin{figure}
  \begin{minipage}[t]{0.48\textwidth}
    \includegraphics[width=\textwidth]{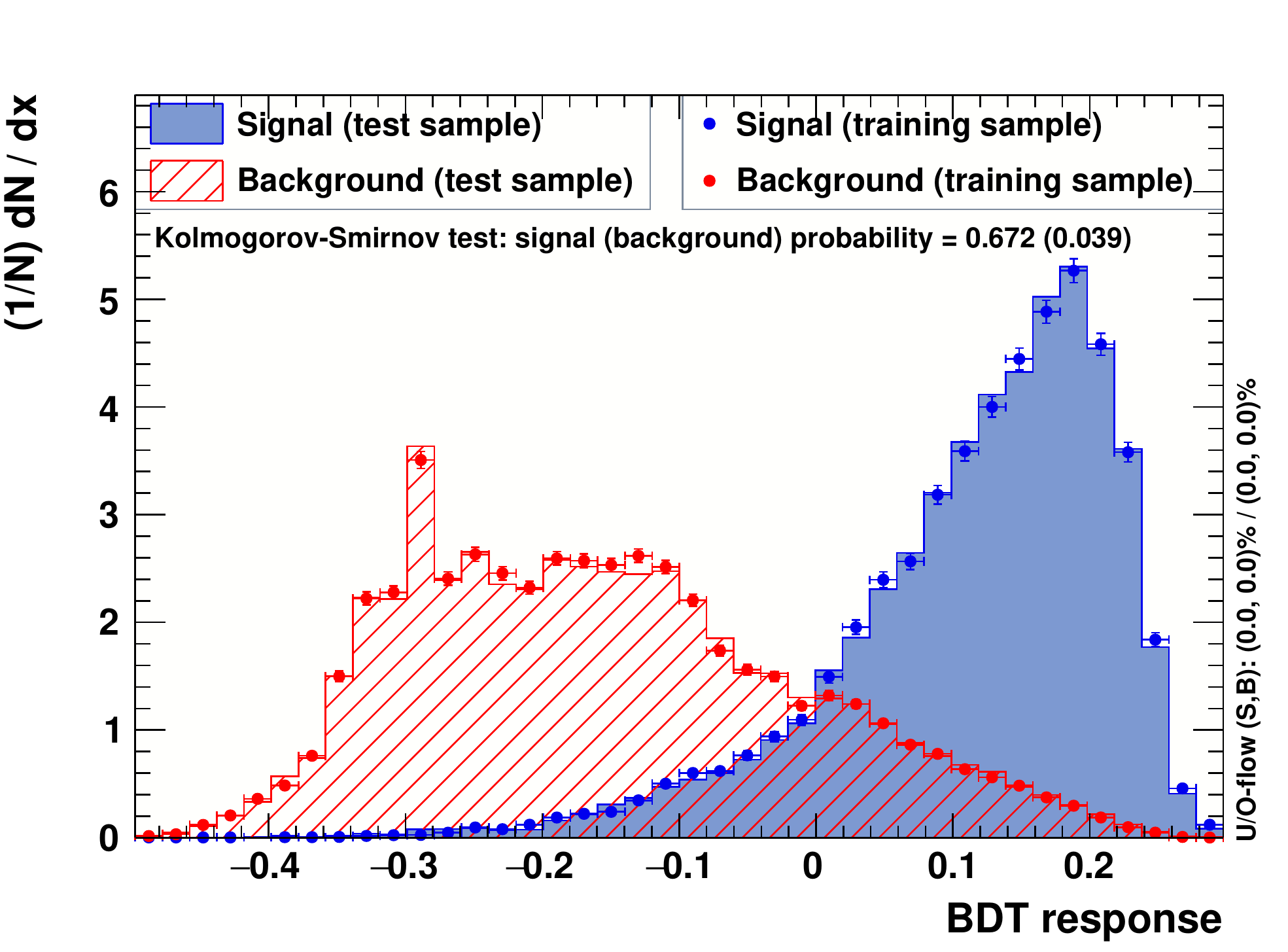}
  \cliclabel{6.15cm}{1.7cm}\vspace*{-5mm}
    \caption{Distribution of BDT response for the signal (red) and the background (blue) for 380~GeV. Points stand for training samples and plain histograms for test samples.}
    \label{fig:results}
  \end{minipage}%
  \hfill 
  \begin{minipage}[t]{0.48\textwidth}
    \includegraphics[width=\textwidth]{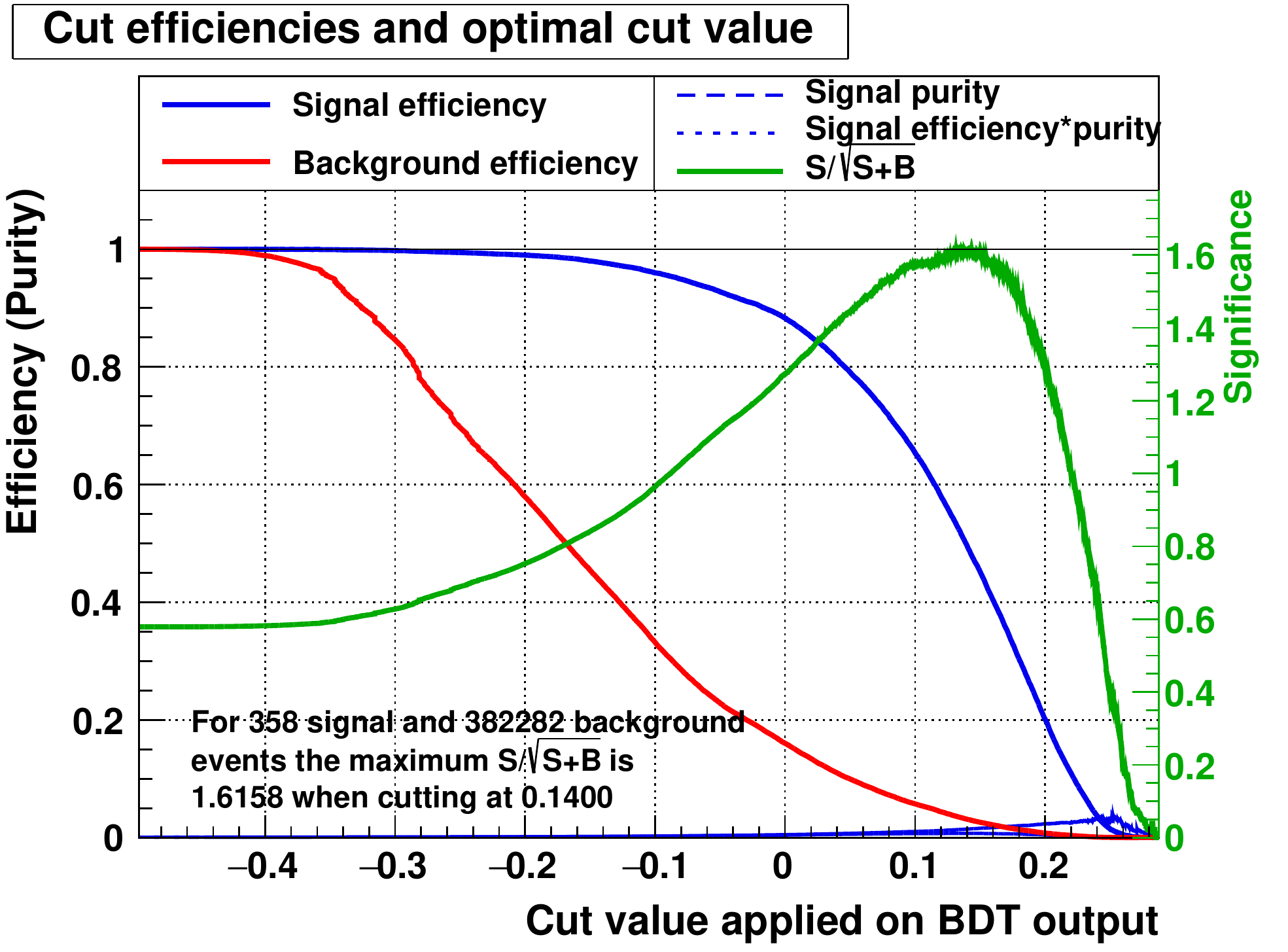}
  \cliclabel{6.25cm}{6.7cm}\vspace*{-5mm}
    \caption{Significance for 380~GeV depending on BDT cut (green) assuming that only 1\% of Higgs bosons decay invisibly.}
    \label{fig:significance}
  \end{minipage}
\end{figure}

The distributions of these variables for the signal and the background
samples are shown in Figure \ref{fig:variables}.
Various sets and numbers of parameters were tested; the above choice
was selected as optimal, resulting in the most efficient event
classification.
Distributions of the BDT algorithm response for considered signal
and background event samples (after preselection cuts) are shown in Figure
\ref{fig:results}.
Consistent distributions obtained for training and test event samples
confirm proper optimization of the BDT algorithm (no overtraining).
Most of the background events can be easily distinguished from the
signal, but there is also a significant contribution of background
events for which BDT response values are positive, consistent with the
response expected for signal events.
This indicates that it is not possible to achieve full separation
between the signal and the background processes.
One should note that about 0.1\% of SM Higgs boson decays result in
fact in the invisible final state ($\PH \to \PZ\PZ^* \to
\PGn\PAGn\PGn\PAGn$), which is included in the background simulation. 

In the final step of the analysis, we select the cut on the BDT
response which gives the highest significance for the expected
signal.
The dependence of the signal significance at 380~GeV on the BDT algorithm
response cut is shown in Figure \ref{fig:significance}, 
for the signal sample normalised to  BR$(\PH \to inv)=1\%$.
The highest significance for invisible Higgs decays at 380\,GeV CLIC is
obtained for a BDT response cut of about 0.14, corresponding to a
BDT selection efficiency for signal events of about 50\% and background
rejection efficiency of about 95\%. 

\begin{figure}
  \begin{subfigure}[t]{0.48\textwidth}
    \includegraphics[width=\textwidth]{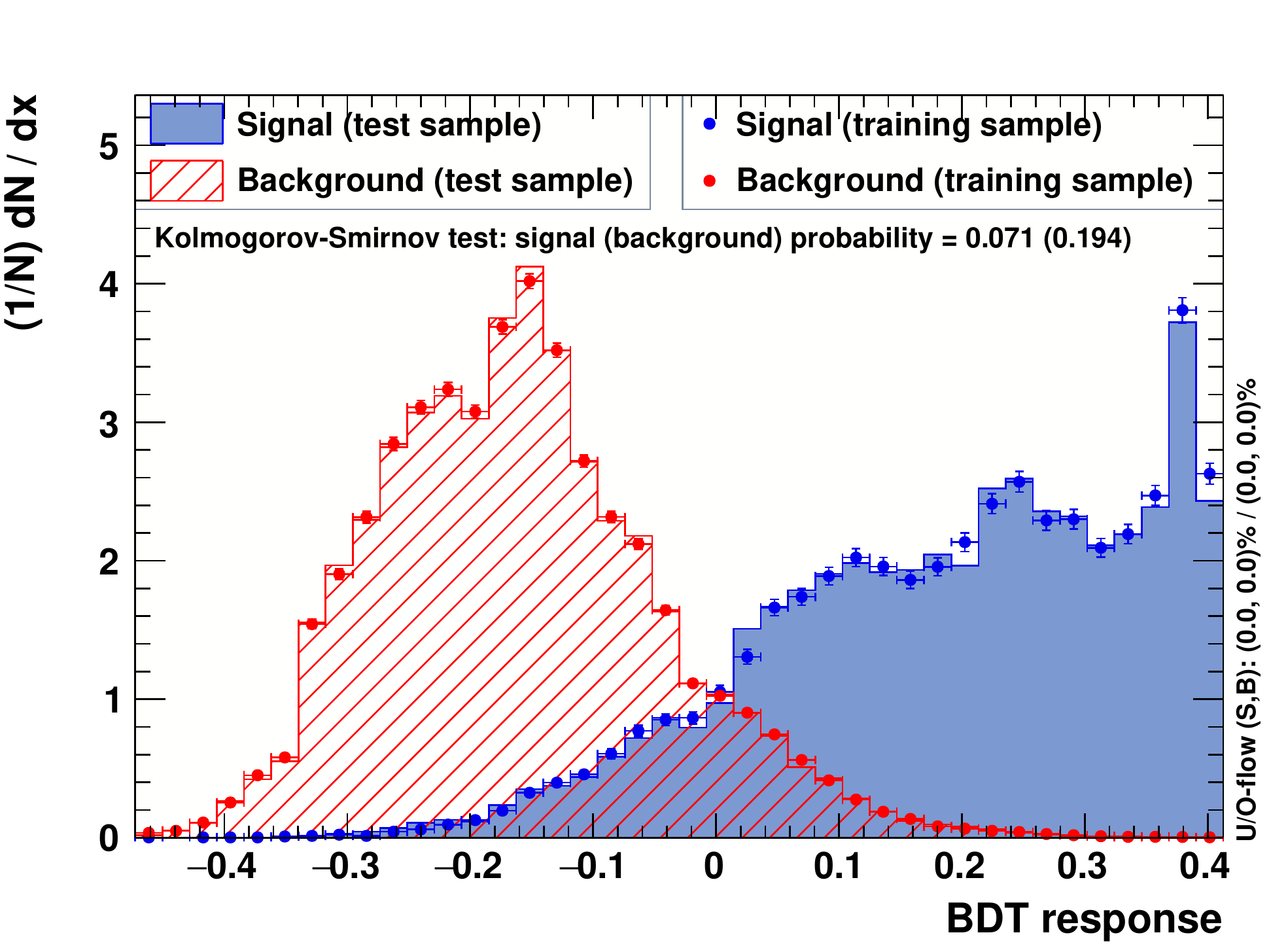}
    \caption{negative polarisation}
    \label{fig:resneg}
  \end{subfigure}%
  \hfill 
  \begin{subfigure}[t]{0.48\textwidth}
    \includegraphics[width=\textwidth]{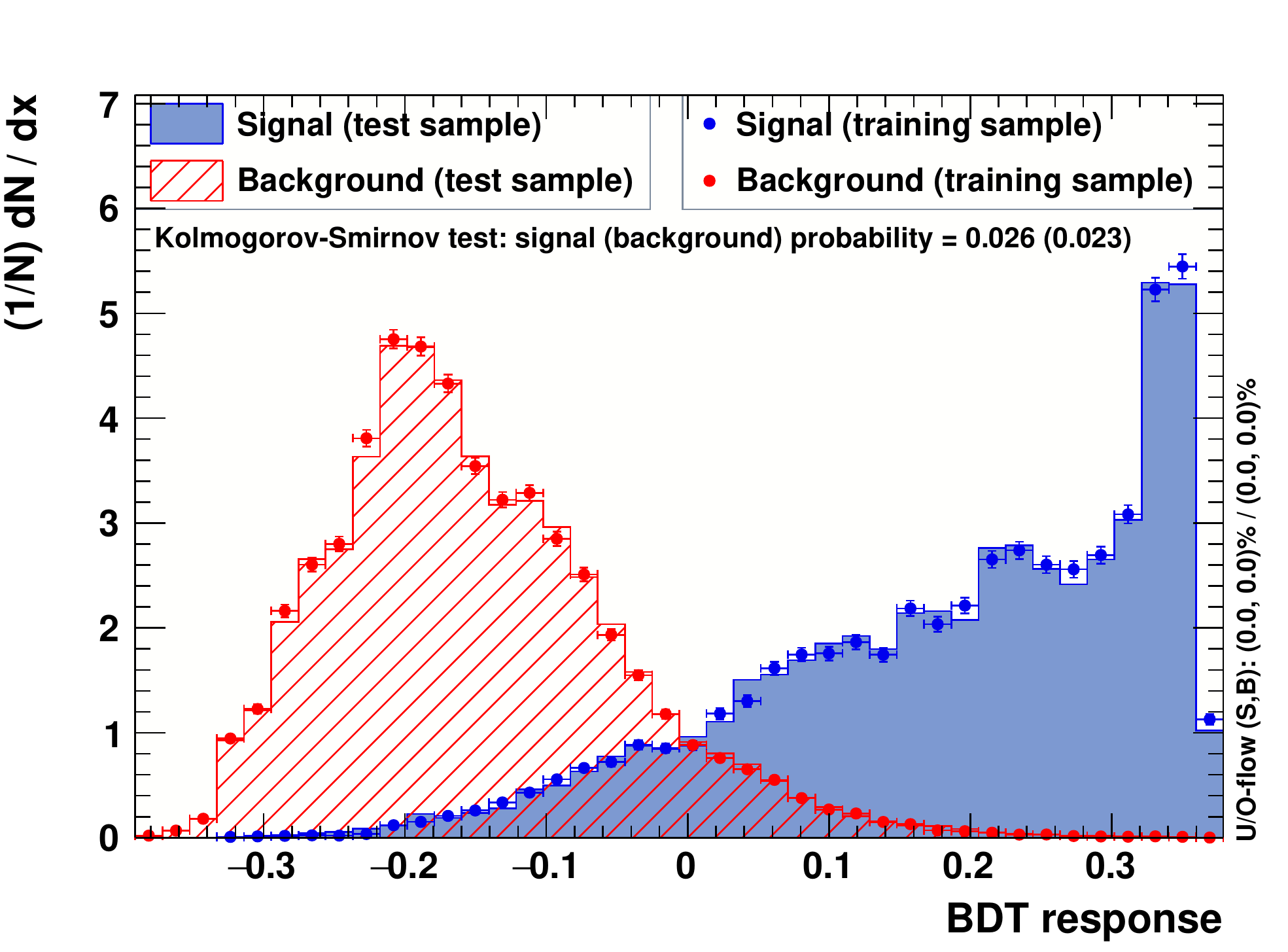}
    \caption{positive polarisation}
    \label{fig:respos}
  \end{subfigure}
  \cliclabel{6.05cm}{1.7cm}
  \cliclabel{6.05cm}{9.9cm}
  \caption{Distribution of BDT response for the signal (red) and the
    background (blue) for CLIC running at 1.5\,TeV, with electron beam
    polarisation of $-80\%$ (left) and $+80\%$ (right).
    Points stand for training samples and plain histograms for test
    samples.}
  \label{fig:results15}
\end{figure}

The same analysis procedure was applied for signal and background samples
generated for CLIC running at 1.5\,TeV, separately for two considered
electron beam polarisation settings.
Distributions of the BDT algorithm response for considered signal
and background event samples (after preselection cuts) are shown in Figure
\ref{fig:results15}.
A similar level of signal-background separation is obtained for each
polarisation. 

\section{Results}
As a result of the preselection and selection procedures presented in
the previous section, expected numbers of background events and
efficiency of signal event selection are obtained.
These can be translated into a constraint on the invisible branching
ratio of the 125\,GeV Higgs boson.
For the first stage of the CLIC accelerator, assuming that the
measured event distributions are consistent with the predictions of the
Standard Model, the expected 95\% CL limit\footnote{Assuming that
  one-side confidence limit of 95\% corresponds to significance of
  1.64.} is: 
\begin{center}
BR$(\PH \to inv)<1.0\%$.
\end{center}
A significance above $5 \sigma$, necessary to confirm the discovery
of a new decay channel (and therefore also existence of new, invisible
particles), is expected for an invisible Higgs boson branching ratio
above 3.1\%.
Presented results, based on the fast simulation of the CLICdet
detector model, are consistent with results of the previous
study~\cite{pop}, BR$(H\to inv) < 0.94\%$,  based on the full detector
simulation, when the difference in the included background channels, collision energy and integrated luminosity is taken into account.\footnote{For collision energy of
  350~GeV, assumed in~\cite{pop}, the expected cross section for
  $\Pep\Pem\to\PQq\PAQq\PH$ is 93~fb, as compared to 82~fb at 380~GeV
  (see Table~\ref{tab1}). In the previous analysis, beamstrahlung and EPA photon interactions were not included. The decrease of the cross section and the increase of the background events  is compensated by factor 2 increase in the assumed integrated luminosity.}

The analysis procedure described above, developed to discriminate between
the background of different SM processes and the signal of invisible
Higgs boson decays was also used to estimate the expected sensitivity
of CLIC experiment to production and invisible decays of a new scalar
state.
Signal simulation was based on the Standard Model implementation in
\whizard (SM\_CKM model), with the modified Higgs boson mass and width
(according to SM predictions for given mass).
Samples of events with production of the new, hypothetical scalar particle
$\PH'$ with the emission of two quarks ($\Pep\Pem\to \PH'\PZ \to
inv+\PQq\PQq$) were generated for masses of the new scalar in the
range 120-280~GeV (for the first stage of CLIC) and
150-1200~GeV (for the second stage).
As for invisible SM Higgs boson decays, the produced
particle $\PH'$ was defined as stable in \whizard and \pythia, and non-interacting in \delphes, to model its invisible decays. 
In this configuration,  $\PH'$ is always produced in \whizard with the
assumed mass, corresponding to the narrow-width approximation. 

\begin{figure}
\begin{center}
\begin{subfigure}[b]{0.48\textwidth}
  \includegraphics[width=\textwidth]{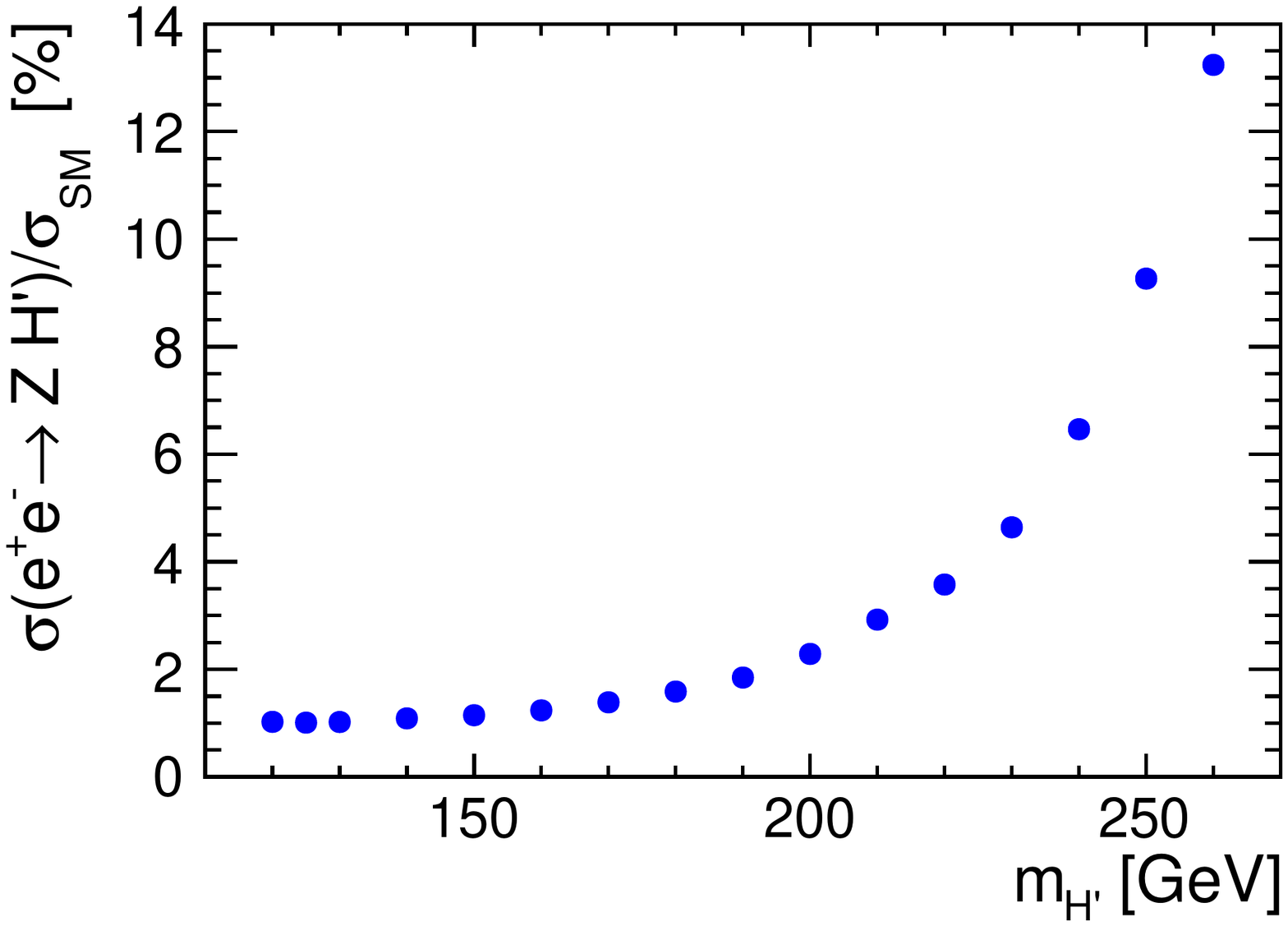}
  \caption{380~GeV}
  \label{fig:limit1}
  \end{subfigure}
  \begin{subfigure}[b]{0.48\textwidth}
    \includegraphics[width=\textwidth]{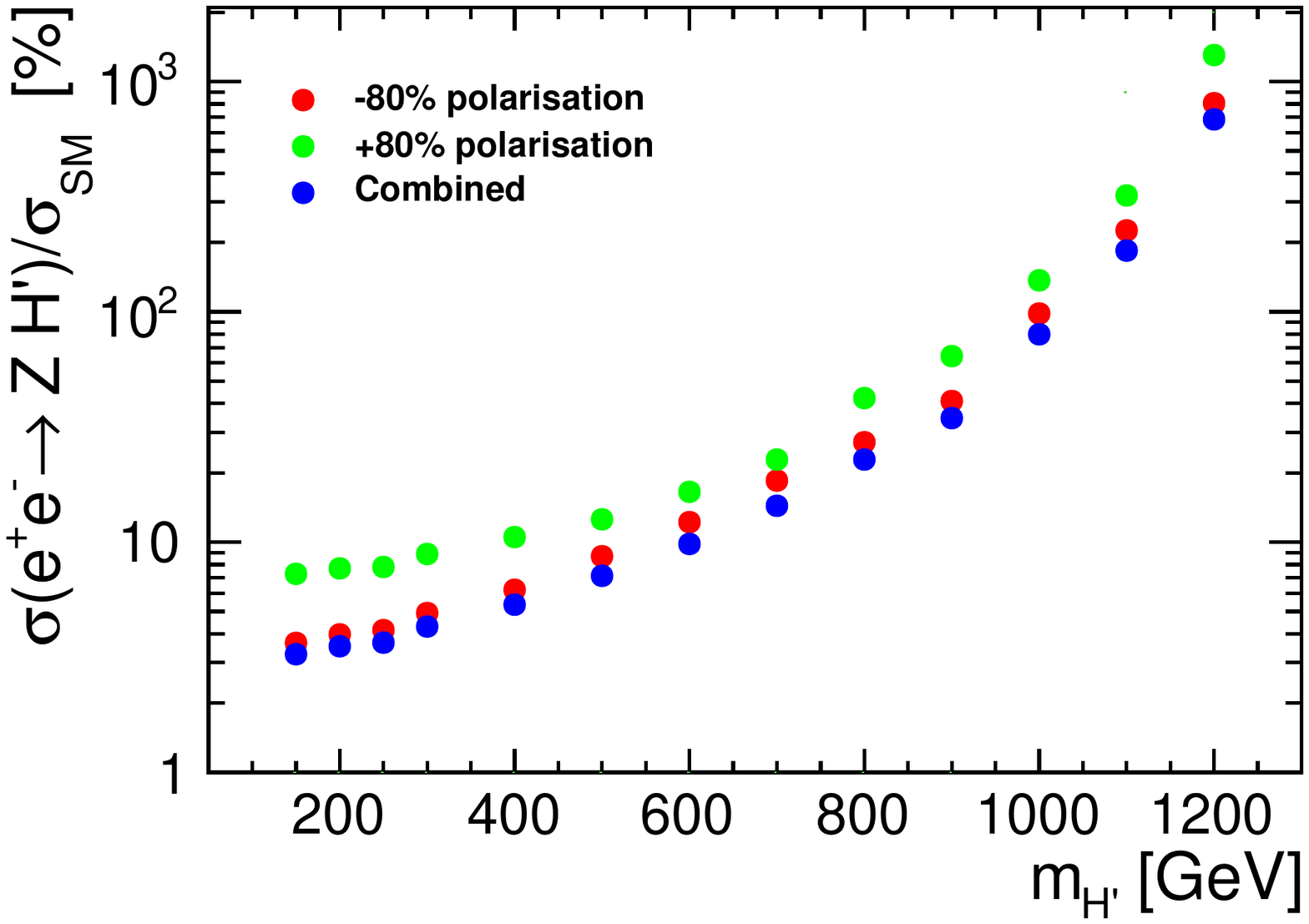}
  \caption{1.5~TeV}
  \label{fig:limit2}
  \end{subfigure}
  \cliclabel{6.4cm}{-10.9cm}
  \cliclabel{6.4cm}{4.5cm}
    \caption{Expected limits on the production cross section of the
      new scalar $\PH'$, relative to the expected SM Higgs production
      cross section, as a function of its mass, for CLIC running at
      380\,GeV (left) and 1.5\,TeV (right). New scalar is assumed to
      have only invisible decay channels, BR$(\PH'\to inv) = 100\%$. } 
    \label{fig:limit}
  \end{center}
\end{figure}

After application of the same preselection cuts, we used the BDT
algorithm, with the same set of input variables, for the final signal event selection.
The algorithm was trained separately for each considered scalar mass 
(each generated signal sample).
With the BDT response cut optimised for signal significance, we
extract an expected limit on the cross-section for the production of
the new scalar $\PH'$ as a function of its mass, assuming its
invisible decays, BR$(H'\to inv) = 100\%$. 
Obtained limits at 95\% CL,  
relative to the expected cross section for the production of the
SM-Higgs boson (for given mass), are presented  
as a function of the assumed scalar mass in Figures
\ref{fig:limit1} and \ref{fig:limit2} for 380\,GeV and 1.5\,TeV, respectively.

The results indicate that the experiment at CLIC will be able to
exclude new scalar production with rate of about 1\% of the SM
production cross-section for masses up to about 170 GeV 
(assuming only invisible decay channels).
For higher masses the experimental sensitivity decreases mainly due to
the decreasing production cross section.

For the second CLIC stage sensitivity to production and invisible
decays of the light Higgs-like scalars is smaller than at 380\,GeV,
mainly due to the decreasing signal cross section and higher
background levels.
The expected limit on the invisible decays of SM Higgs boson is about
3\%.
Assuming the production cross section given by the SM predictions,
the second stage of CLIC will be sensitive to the new `invisible'
scalars up to about 1\,TeV.

\section{Interpretation}

\begin{figure}
    \includegraphics[width=0.48\textwidth]{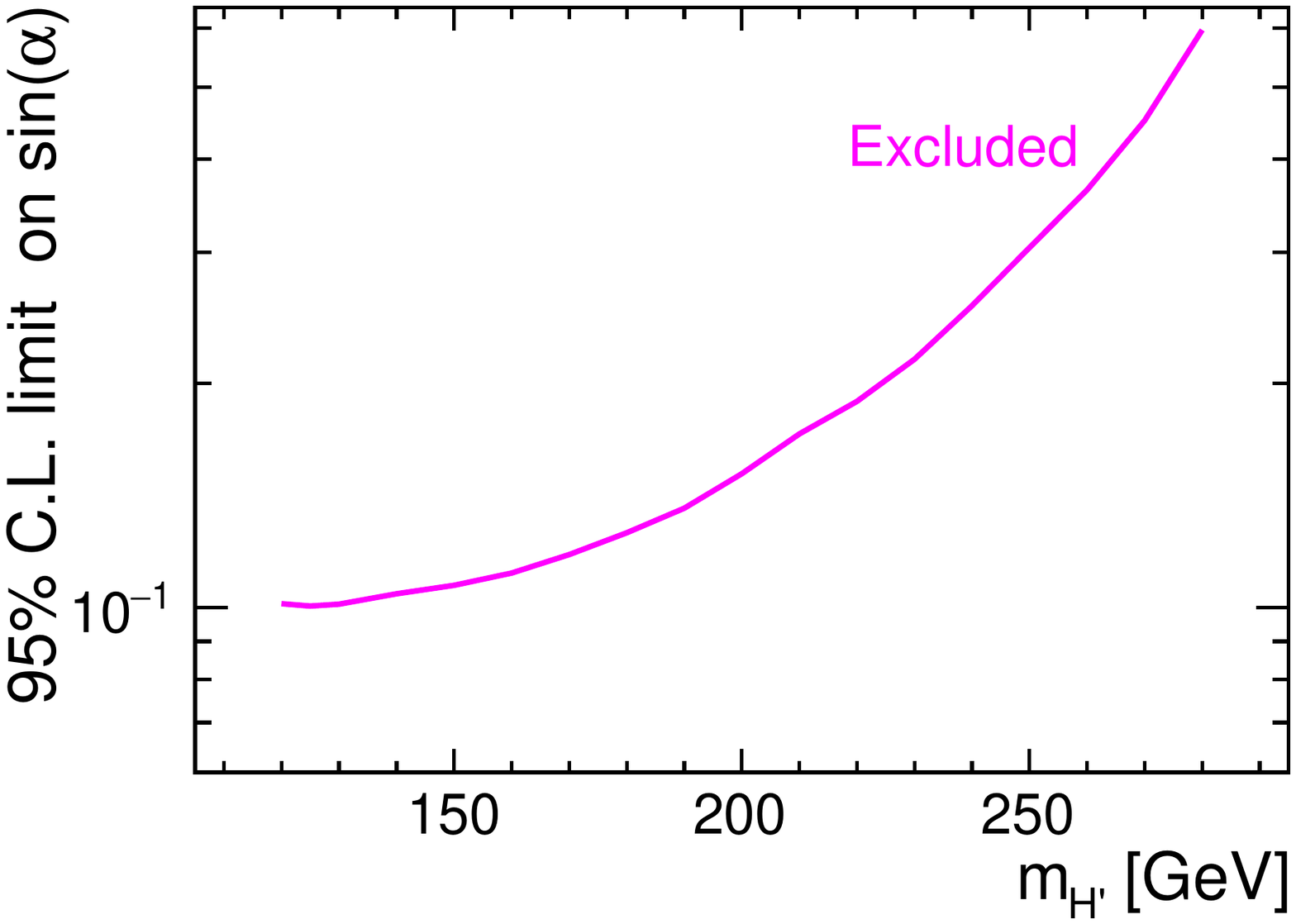}
  \hfill 
  \includegraphics[width=0.48\textwidth]{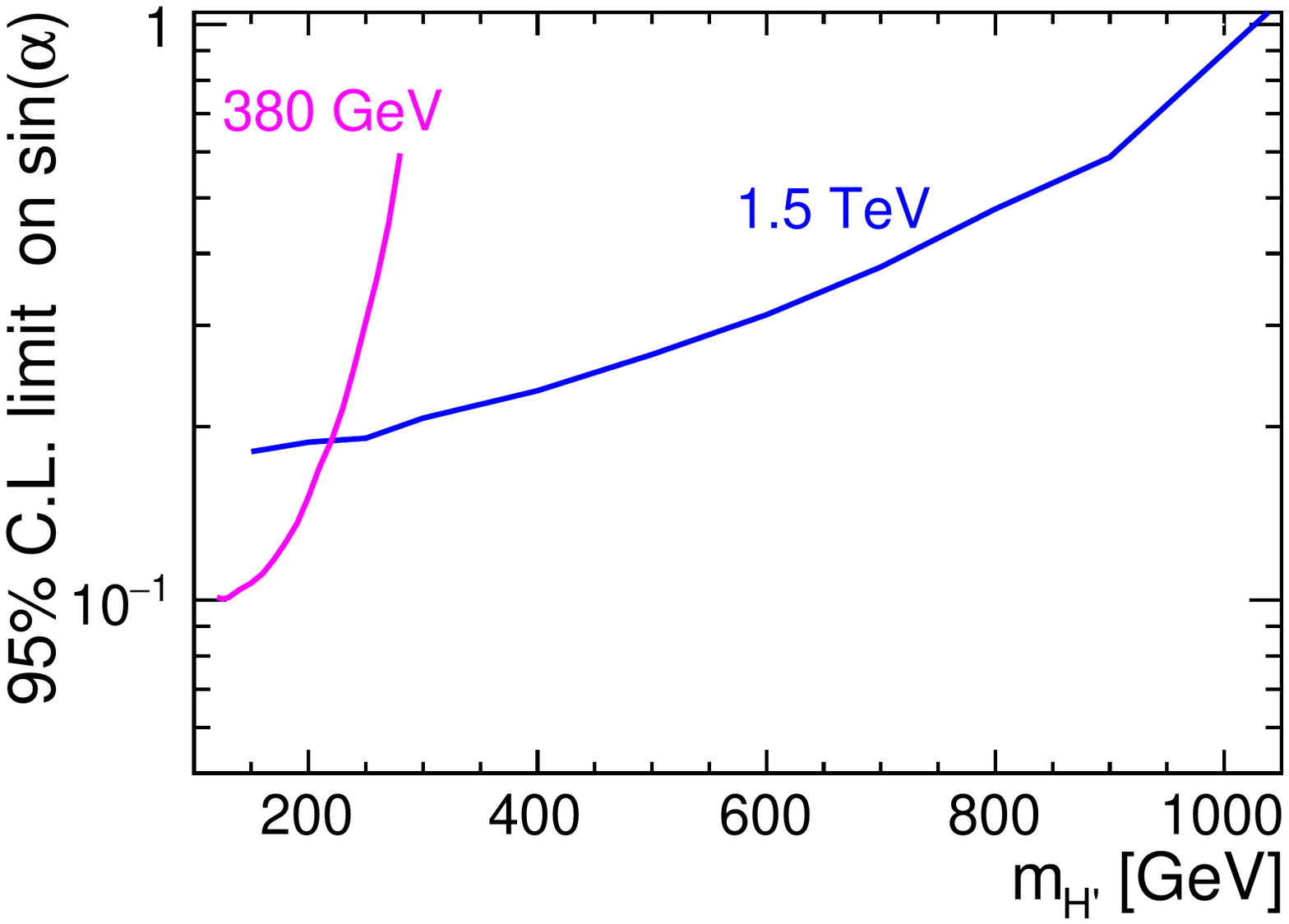}
  \cliclabel{5.75cm}{2.15cm}
  \cliclabel{5.75cm}{10.3cm}
      \caption{Expected limits on the sine of the scalar sector mixing
        angle, $\sin\alpha$, as a function of the $\PH'$ mass, for
        CLIC running at 380\,GeV (left and right plots) and
        1.5\,TeV (right plot).}
          \label{fig:excluded}
\end{figure}

The expected limits on invisible decays of the 125\,GeV Higgs boson
and limits on the production of new "invisible" scalars, which were
obtained in a model-independent approach, can also be used to
constrain different BSM scenarios.
As an example, we demonstrate the possibility of constraining
parameters of the VFDM model~\cite{vfdm1,vfdm2}.
The Standard Model (SM) is extended by the spontaneously broken extra
$U(1)_X$ gauge symmetry and a Dirac fermion.
To generate mass for the dark vector $X_\mu$ the Higgs mechanism with
a complex singlet $S$ is employed in the dark sector.
Dark matter candidates are the massive vector boson $X_\mu$ and two
Majorana fermions $\psi_\pm$.
The spontaneous symmetry breaking in the dark sector results in an
additional scalar state $\phi$.
This state can mix with the SM Higgs field $h$ implying existence of
two mass eigenstates: 
\begin{equation} 
 \left( 
\begin{array}{c}
\PH\\ 
\PH'
\end{array} 
\right) =
\left( 
\begin{array}{cc}
\cos \alpha   & \sin \alpha \\ 
-\sin \alpha &\cos \alpha
\end{array} 
\right)
\left(
\begin{array}{c}
h\\ 
\phi
\end{array} 
\right) \; ,
\nonumber
\end{equation} 
where we assume that $\PH$ is the observed 125\,GeV state. 
If $\alpha \ll 1$, it is SM-like, but it can also decay invisibly (to
dark sector particles) via the $\phi$ component (BR$(\PH\to inv)\sim \sin^2\alpha$).  
If $\PH'$ is also light, it can be produced in e$^+$e$^-$ collisions
in the same way as the SM-like Higgs boson.
We assume in the following that invisible decays to dark matter sector
particles dominate for $\PH'$ (BR$(\PH'\to inv) \approx 100\%$). 
If this is the case,\footnote{This assumption is a good approximation
  in a wide range of model parameters, although a small part of the new
  scalar decays must also be visible because otherwise the scalar
  would not be produced.}
the cross section for new scalar production
corresponding to the limits presented in the previous section can be
writen as:
\begin{equation}
\sigma_{H'} = \sigma^{m_H = m_{H'}}_{SM} \cdot \sin^{2}(\alpha) \; ,
\nonumber
\end{equation}
where $\sigma_{H'}$ is the cross-section for the production of a new
scalar of mass $m_{H'}$, and $\sigma^{m_H = m_{H'}}_{SM}$ is the
cross-section for the production of the Higgs boson in the Standard
Model with the same mass value.
Limits on the sine of the mixing angle, $\sin\alpha$, resulting from
the cross-section limits presented in Figure \ref{fig:limit} are shown
in Figure \ref{fig:excluded}. 

\begin{figure}
    \centering
    \includegraphics[width=0.5\textwidth]{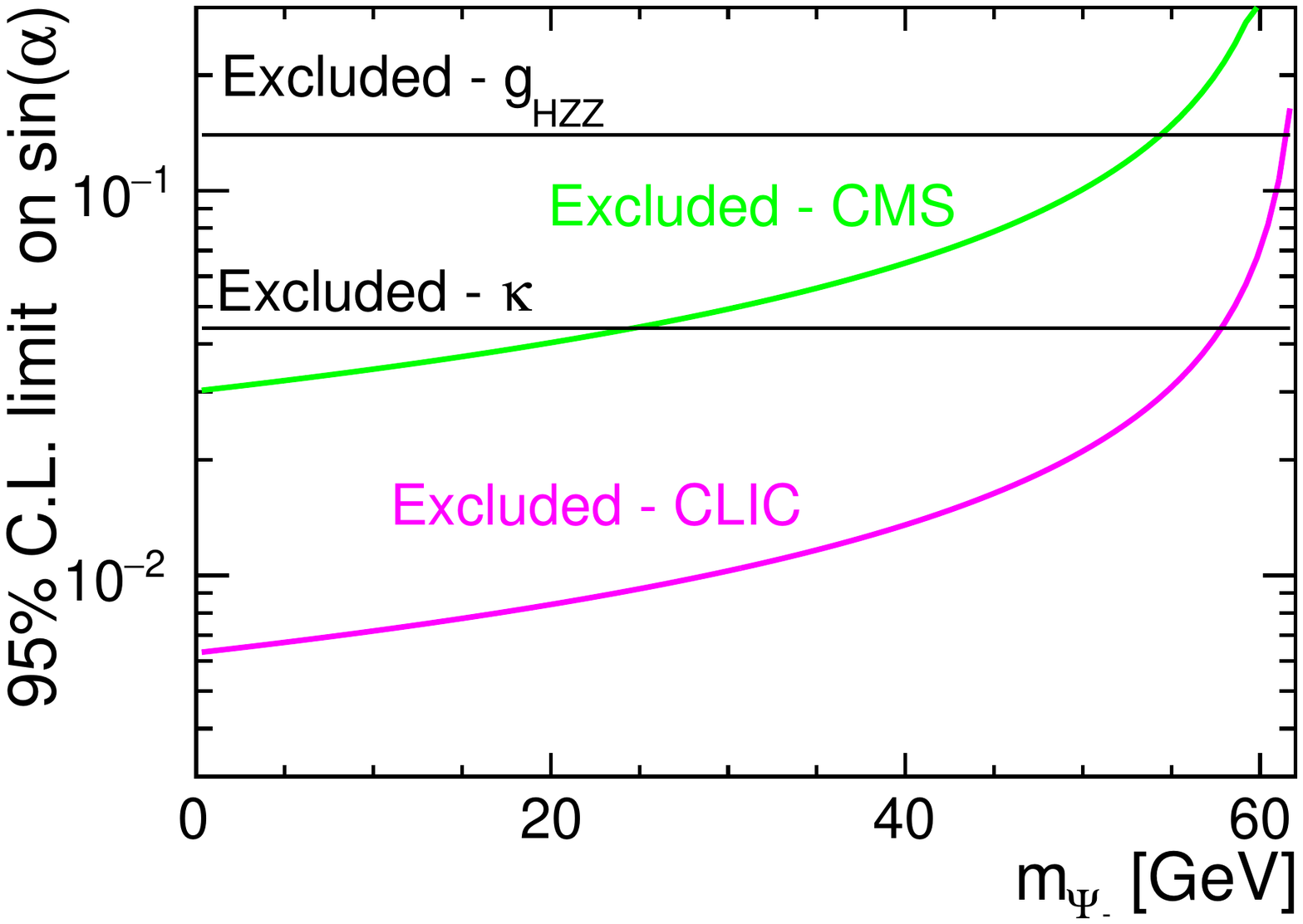}
    \cliclabel{5.95cm}{-3.4cm}
    \caption{Expected limit on the sine of the scalar sector mixing
        angle, $\sin\alpha$, as a function of the $\Psi_{-}$ particle
        mass of the VFDM model, for CLIC running at
        380\,GeV. Indicated by the green curve is the limit
        corresponding to the current constraint, BR$(\PH\to inv) < 19\%$,
        from the CMS experiment\cite{CMS}. The horizontal line indicates
        the indirect limit expected from the measurement of
        the $g_{_{HZZ}}$ and $\kappa$ couplings. 
        } 
    \label{fig:excluded2}
\end{figure}

The mixing angle in the VFDM model can also be constrained by
analysing the limit on the invisible branching ratio for the SM-like Higgs
boson BR$(\PH\to inv)$. 
When the contribution of the H'H' decay channel can be neglected,
the invisible partial width of the Higgs boson, $\Gamma_{inv}$, is
proportional to $\sin^{2}(\alpha)$, but depends also on other model
parameters, in particular on the dark sector coupling constant,
$g_{x}$, the mass of the vector dark matter, $m_X$, and the masses of
the fermionic dark matter particles, $m_{\Psi_{-}}$ and
$m_{\Psi_{+}}$.\footnote{The two fermionic states are defined in such
  a way that $m_{\Psi_{-}}  \le  m_{\Psi_{+}}$.}  
Constraints on the scalar sector mixing angle, resulting from the
limits on the invisible decays of the SM-like Higgs boson expected at
380\,GeV CLIC, are shown in Figure \ref{fig:excluded2}.
Expected limits on $\sin\alpha$, plotted as a function of the
$\Psi_{-}$ particle mass, are based on the invisible decay
widths\footnote{The scaling of SM (visible) decay width with factor
  cos$^{2}$($\alpha$) was  neglected for the considered range of
  $\sin\alpha$.} 
calculated with \whizard,
for $g_x$ = 1 and $m_X$ = $m_{\Psi_{+}}$ = 200 GeV.
Also indicated in the Figure~\ref{fig:excluded2} are the indirect
limits on the mixing angle, which can be set at CLIC from the
analysis of the Higgs coupling measurements.
Due to the mixing with the $\phi$ state, all couplings of the SM-like
Higgs boson to SM particles are scaled by $\cos(\alpha)$.
In particular, the coupling of the Higgs boson to the $\PZ$ bosons,
$g_{_{HZZ}}$, is given by:
\begin{equation}
  g_{_{HZZ}} = g_{_{HZZ}}^{_{(SM)}} \cdot \cos(\alpha)\;.
\nonumber
\end{equation}
It is expected that the experiment at 380\,GeV CLIC will be able to
measure $g_{_{HZZ}}$ in a model-independent approach with an accuracy of
0.6\%~\cite{gHZZ}. 
If no deviations from SM are observed, the corresponding 95\% C.L. 
limit on the mixing angle in the VFDM model is: 
\begin{equation}
|\sin(\alpha)| < 0.14\;.
\nonumber
\end{equation}
If the Higgs coupling fit is performed with the
assumption that the Higgs boson couplings to all SM particles 
scale by the same factor, $\kappa$, much stronger constraints 
can be set \cite{deBlas:2018mhx}.
After three CLIC running stages, the overall scaling of the Higgs boson
couplings shoul be known to
\begin{equation}
 \Delta \kappa = 0.06\,\% \, .
\nonumber
\end{equation}
This corresponds to 95\% C.L. 
limit on the mixing angle in the VFDM model of: 
\begin{equation}
|\sin(\alpha)| < 0.044\;.
\nonumber
\end{equation}
Also indicated is  Figure~\ref{fig:excluded2} is the
limit on invisible Higgs boson decays from the CMS experiment,
BR$(\PH\to inv) < 19\%$~\cite{CMS}. 
For masses of dark matter particles up to about 60\,GeV, CLIC will
allow to set much better constraints on the mixing angle in the scalar
sector than it will be possible with indirect methods.  

\section{Summary}

We studied the possibility of measuring invisible Higgs boson decays
with CLIC running at 380 GeV and 1.5 TeV, taking background 
contributions from photon-photon and electron-photon interactions into 
account.
The analysis is based on the \whizard event generation and
fast simulation of the CLIC detector response with \delphes.
An approach consisting of a two step analysis was used to optimize
separation between signal and background processes.
The expected limit on the branching ratio for invisible Higgs boson decays
is BR$(H \to inv) < 1.01\%$.
This result is consistent with the previous analysis carried out with
full simulation of the detector response.
This shows that the fast simulation method used in the presented
analysis should also be applicable to studies concerning new physics
scenarios at CLIC.
The branching-ratio and cross-section limits obtained in the
model-independent approach  can also be used to set limits on the
different extensions of the Standard Model.
In particular, constraints at the percent level can be set on the
scalar sector mixing angle in ``Higgs-portal'' models.

\subsection*{Acknowledgements}

The work was carried out in the framework of the CLIC detector and
physics (CLICdp) collaboration.
We thank collaboration members for fruitful discussions, valuable
comments and suggestions.

The work was partially supported by the National Science Centre
(Poland) under OPUS research projects no 2017/25/B/ST2/00496
(2018-2021) and 2017/25/B/ST2/00191, and a HARMONIA project under 
contract UMO-2015/18/M/ST2/00518 (2016-2019).

\printbibliography[title=References]

\end{document}